\documentclass[aps, twocolumn, groupedaddress,superscriptaddress, showpacs,pre,nofootinbib]{revtex4-1}
\usepackage{graphicx}
\usepackage[title]{appendix}
\usepackage{array}
\usepackage{setspace,transparent}
\usepackage{color}

\usepackage{fontenc,subcaption,ragged2e,amsmath}
\usepackage{mathtools,amssymb,amssymb,nicefrac,tikz}
\usepackage{soul}
\usepackage[font=small,labelfont=bf]{caption}
\usepackage[utf8]{inputenc}

\DeclareCaptionJustification{justified}{\justifying}
\newcommand{\RN}[1]{%
  \textup{\uppercase\expandafter{\romannumeral#1}}%
}
\DeclareMathAlphabet{\mathpzc}{OT1}{pzc}{m}{it}
\DeclareCaptionJustification{justified}{\justifying}
\begin{document}

\title{Microscopic intervention yields abrupt transition in interdependent magnetic networks}
\author{Bnaya Gross}
\thanks{Corresponding author: bnaya.gross@gmail.com}
\affiliation{Department of Physics, Bar-Ilan University, 52900 Ramat-Gan, Israel}
\author{Ivan Bonamassa}
\affiliation{Department of Network and Data Science, CEU, Quellenstrasse 51, A-1100 Vienna, Austria}
\author{Shlomo Havlin}
\affiliation{Department of Physics, Bar-Ilan University, 52900 Ramat-Gan, Israel}
\date{\today}

\begin{abstract}
    The study of interdependent networks has recently experienced a boost with the development of experimentally testable materials that physically realize their critical behaviors, calling for systematic studies that go beyond the percolation paradigm. Here we study the critical phase transition of interdependent spatial magnetic networks model where dependency couplings between networks are realized by a thermal interaction having a tunable spatial range. We show how the critical phenomena and the phase diagram of this realistic model are highly affected by the range of thermal dissipation and how the latter changes the transition from continuous to abrupt. Furthermore, we show that microscopic interventions of localized heating and localized magnetic field yields a macroscopic phase transition and novel phase diagrams. Our results provide novel and realistic insights about controlling the macroscopic phases of interdependent materials by means of localized microscopic interventions.
\end{abstract}

\maketitle

In the last few decades, network theory has been proven useful in describing collective phenomena in various fields ranging from social \cite{borgatti2018analyzing}, technology \cite{barabasi-science1999,albert1999diameter}, biology \cite{junker2011analysis,alm2003biological}, and medicine \cite{barabasi2011network}. It gained a significant boost with the development of the interdependent networks paradigm \cite{buldyrev2010catastrophic,parshani2010interdependent}, showing how dependency interactions between macroscopic systems lead to the emergence of novel phenomena that do not occur in their isolated counterparts. The resilience of such networks is usually studied by percolation theory \cite{bunde1991fractals,staufferaharony} which has been extensively applied in the last decade for studying and understanding different structural and functional properties of isolated and interacting networks. ~\cite{gao2012networks,wei-prl2012,gao2011robustness,boccaletti2014structure,stippinger2014enhancing,lee2017universal,kivela2014multilayer,majhi2022dynamics,battiston2018multiplex,gross2021interdependent,morris2012transport}.  
\par
Despite the extensive study of percolation of abstract interdependent networks, applying the interdependent paradigm in physical networks as \textit{physical interdependent networks} (PINs) is a challenge that has remained, so far, unexplored. Some steps forward happened recently with the theoretical study of interdependence as thermal couplings in random magnetic networks \cite{bonamassa2021interdependent} followed by the first experimental realization of interdependent superconducting networks \cite{bonamassa2022superconductors}.
\par
While most studies in interdependent networks focus on random dependency couplings, only a few of them have focused on the effects that a spatial dependency range has on the model's behaviors. This is, for example, the case of percolation in spatial interdependent networks, where a finite range of dependency links has been proven crucial for exploring the variety of critical processes occurring in these models \cite{wei-prl2012,gross2022fractal,berezin2015localized,vaknin2017spreading}. Thus, despite the study of interdependent \textit{random} (non-spatial) magnetic networks \cite{bonamassa2021interdependent}, the realistic case of spatial physical networks has never been explored.
\par
In this Letter, we study a realistic model of spatial interdependent magnetic system and analyze the macroscopic effects induced by localized interventions. We find the different effects that localized heating and a localized magnetic field have on the kinetics of nucleating droplets and characterize the phase diagram as a function of the spatial range of the intervention. Our results generalize localized attacks beyond percolation and provide useful insights for future experimental validation.

\begin{figure}
	\centering
	\begin{tikzpicture}[      
	every node/.style={anchor=north east,inner sep=0pt},
	x=1mm, y=1mm,
	]   
	\node (fig1) at (0,0)
	{\includegraphics[scale=0.47]{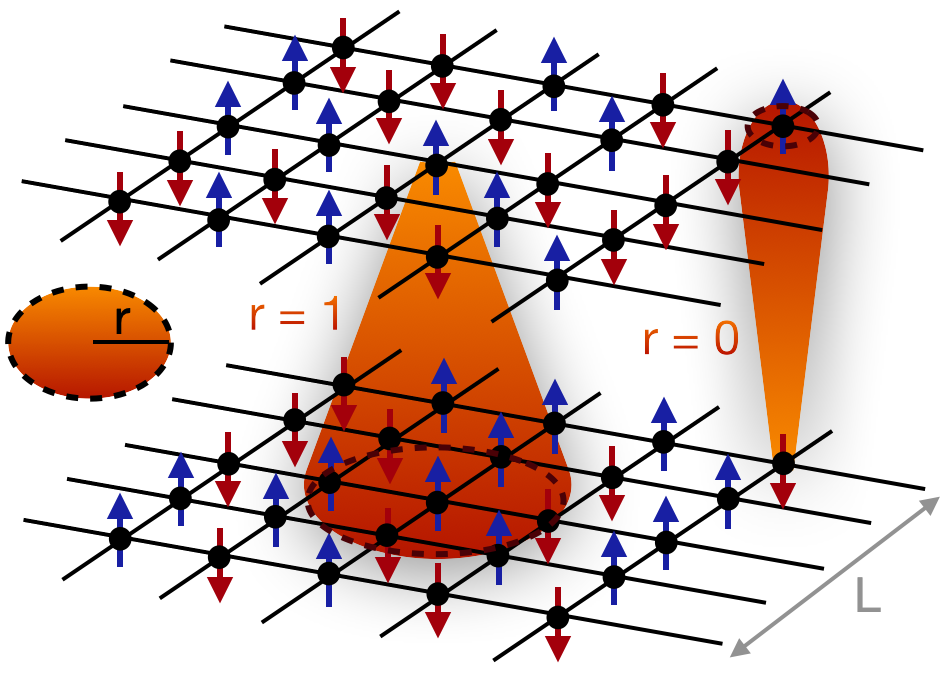}};
	\end{tikzpicture}
	\caption{\textbf{Illustration of the model.} Two $2D$ magnetic layers (lattices) of size $N = L^2$ are interdependent on each other via thermal couplings. The magnetic state of each network $\mu$ is described by their spins configuration $\boldsymbol \sigma_{\mu} = \{ \sigma_1^\mu, \sigma_2^\mu, ...., \sigma_N^\mu \}$ where each node of the network is an Ising spin pointing up (blue arrow) or down (red arrow) having $\sigma_i^\mu = +1, -1$ respectively. Interdependence is realized via thermal coupling where each node in network $\mu$ affects all nodes up to a distance $r$ in the other network $\mu'$ and vice versa.}
	\label{fig:illustration}	
\end{figure}

\begin{figure}
	\centering
	\begin{tikzpicture}[      
	every node/.style={anchor=north east,inner sep=0pt},
	x=1mm, y=1mm,
	]   
	\node (fig1) at (0,0)
	{\includegraphics[scale=0.35]{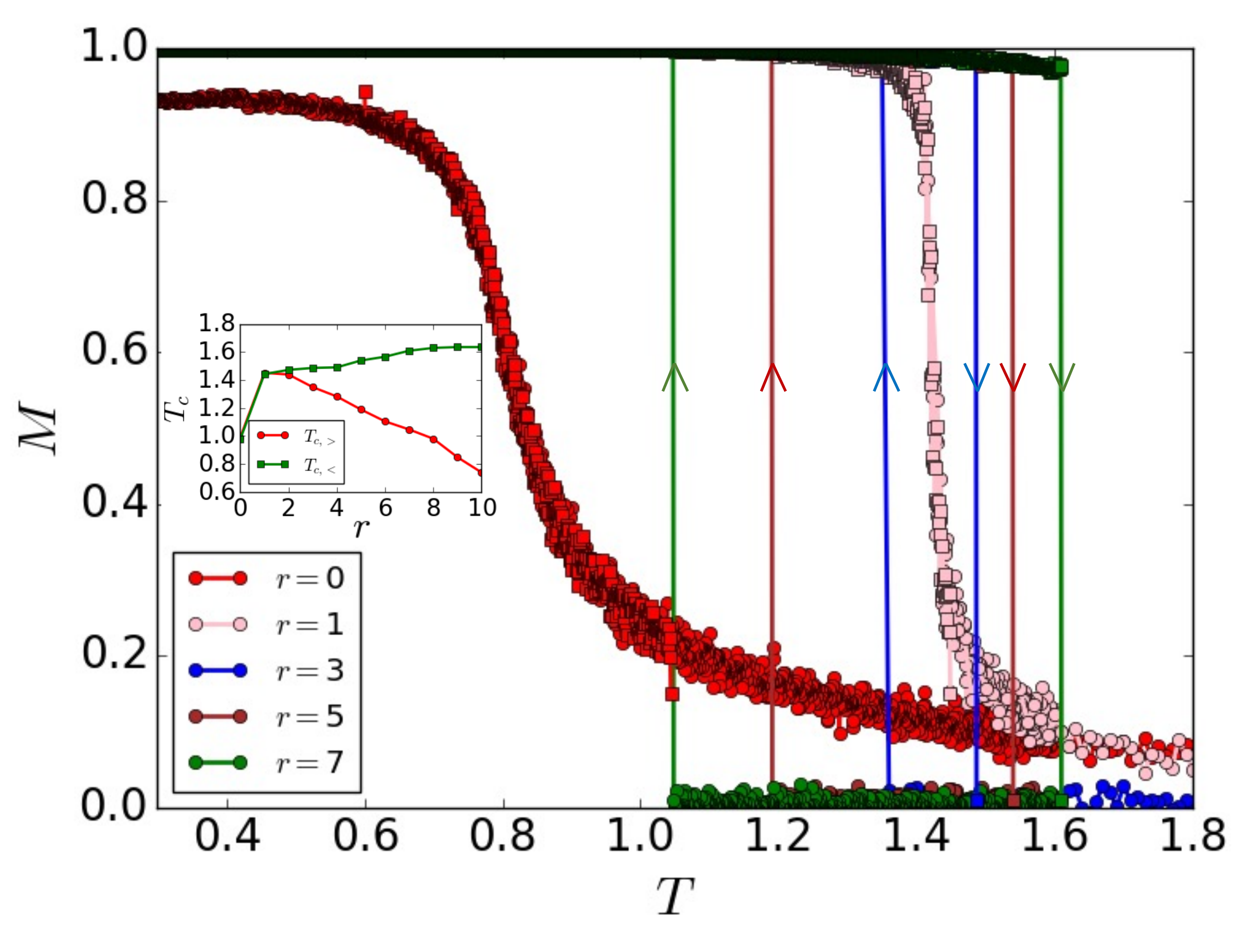}};
	\end{tikzpicture}
	\caption{\textbf{Interdependent magnetization phase transitions.} Magnetization $M$ as a function of temperature $T$ is shown for different values of the dependency interaction range $r$. A smeared-out continuous transition is observed for short interaction range $r < r_c \simeq 2$ and abrupt transition for $r > r_c$. Inset: The critical temperatures, $T_{c,<}(r)$, when increasing temperature (heating) from the ordered state ($M \simeq 1$) to the disordered state, and the critical transitions, $T_{c,>}(r)$ when decreasing temperature (cooling) from the disordered state ($M \simeq 0$) to the ordered state, are the same in the continuous regime $r < r_c$ but are different for $r>r_c$ showing hysteresis. Here $J = 1$, $L = 200$ and performing $10^4$ MCSs.}
	\label{fig:M_T}	
\end{figure}
\begin{figure}
	\centering
	\begin{tikzpicture}[      
	every node/.style={anchor=north east,inner sep=0pt},
	x=1mm, y=1mm,
	]   
	\node (fig1) at (0,0)
	{\includegraphics[scale=0.48]{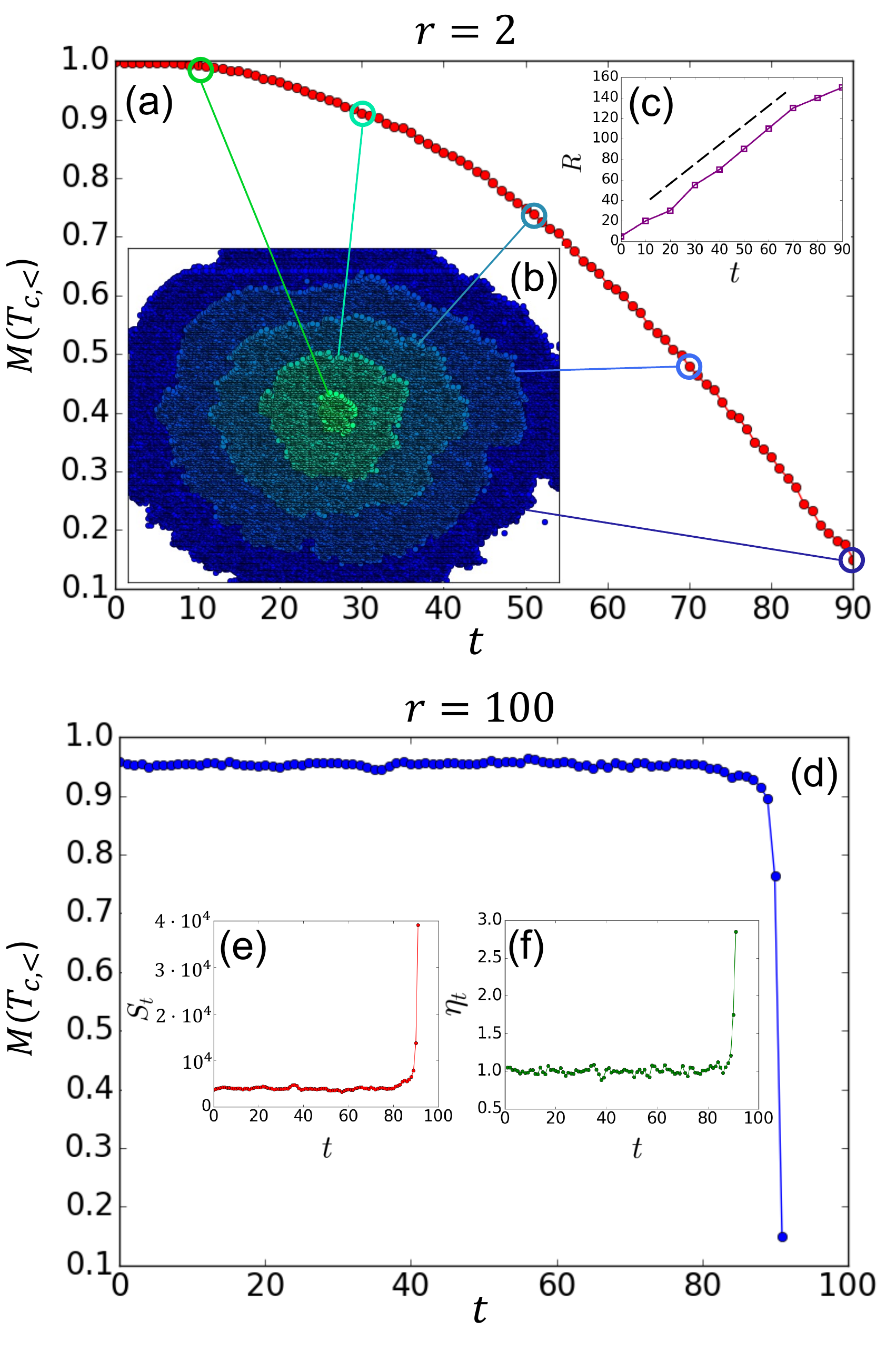}};
	\end{tikzpicture}
	\caption{\textbf{Critical dynamics at $T_{c,<}$. (a)} For low values of $r$ above $r_c$, nucleation transition is observed with a parabolic shape decrease of the magnetization associated with \textbf{(b)} a circular area of disordered spins that spontaneously appears at $T_{c,<}$ and increases in time due to the dependency heat interactions between the layers. \textbf{(c)} The radius of the circle $R$ increases linearly with time. \textbf{(d)} For high values of $r$, such as $r=100$ shown here, a plateau is observed where the magnetization remains nearly constant for a long time. At the end of the plateau, the system converges to the disordered phase exponentially fast. \textbf{(e)} The number of flipped spins as a function of time, $S_t$, is constant during the plateau showing \textbf{(f)} a critical branching factor, $\eta_t = S_{t}/ S_{t-1} \simeq 1$. The analogy to percolation of abstract interdependent networks can be seen in Berezin \textit{et al} \cite{berezin2015localized} and Zhou \textit{et al} \cite{zhou2014simultaneous}.
}
	\label{fig:MTc_t}	
\end{figure}
\begin{figure*}
	\centering
	\begin{tikzpicture}[      
	every node/.style={anchor=north east,inner sep=0pt},
	x=1mm, y=1mm,
	]   
	\node (fig1) at (0,0)
	{\includegraphics[scale=0.53]{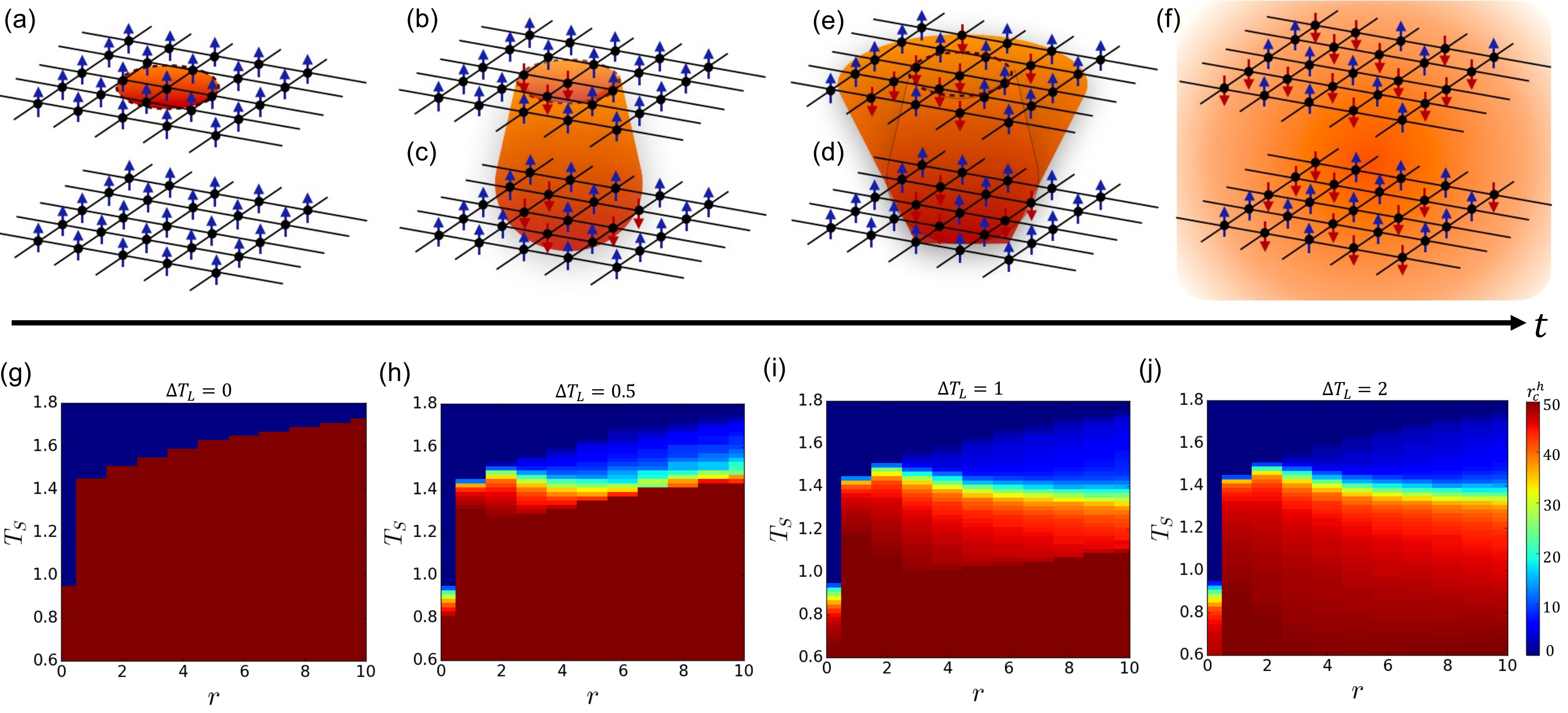}};
	\end{tikzpicture}
	\caption{\textbf{Localized heating. (a)} The upper layer is heated locally within a microscopic radius $r^h$, to temperature $T_S + \Delta T_{L}$. \textbf{(b)} This heating creates a disordered droplet which starts to dissipate heat to the bottom layer. \textbf{(c)} The localized regime in the bottom layer is heated up and becomes disordered as well. \textbf{(d)} The disordered regime in the bottom layer starts to dissipate heat back to the top layer broadening the circle of disorder. \textbf{(e)}  The disordered droplet in the top layer extends due to the dissipation from the bottom layer. \textbf{(f)} This nucleation process continues until the disordered droplet takes over the system. \textbf{Phase diagrams.} While for low values of $r$, but above $r_c$, a spontaneous nucleation transition is observed at $T_{c,<}$, an \textit{induced} nucleation macroscopic phase transition is seen for $T < T_{c,<}$ by an external microscopically localized heating. The phase diagrams of the critical heating radius $r_c^h$ for different heating intensity $\Delta T_{L}$ are shown. \textbf{(g)} For $\Delta T_{L} = 0$ there are no induced nucleation transitions and only two phases separated by $T_{c,<}(r)$ appear, ordered phase (brown) and disordered phase (blue). \textbf{(h)-(j)} Once the system is locally heated, a \textit{metastable regime} appears where localized heating with a critical radius above $r_c^h$ can induce a nucleation transition. The metastable regime expands with the localized heat intensity $\Delta T_{L}$ towards lower temperatures, shrinking the ordered phase. Here $J = 1$ and $L = 100$.}
	\label{fig:phase_diagram_heat}	
\end{figure*}
\underline{\textit{Model.--}}
Let us consider a system composed by two $2D$ magnetic network lattices of size $N = L^2$ placed in a common heat bath of temperature $T$. Each node can represent a ferromagnetic grain and it is endowed with an Ising spin $\sigma = \pm 1$ so that the configuration of spins in network $\mu$ at time $t$ is $\boldsymbol \sigma_{\mu}(t) = \{ \sigma_1^\mu(t), \sigma_2^\mu(t), ...., \sigma_N^\mu(t) \}$, see Fig.~\ref{fig:illustration}. When current is induced in each layer, the networks become thermally coupled~\cite{bonamassa2021interdependent} by a heat dissipation mechanism resulting from the change of local resistance due to electron scattering, similarly to magnetoresistors~\cite{pippard1989magnetoresistance,xiao1992giant}. When spins are locally aligned (ordered), electrons experience weak scattering and the local resistance is low having weak dissipation. On the other hand, when spins are not locally aligned (disordered), strong scattering is expected with high resistance and strong dissipation~\cite{white1992giant}. Since local spins alignment is correlated with resistance which implies heat dissipation, the thermal coupling can be modeled as follows: locally ordered spins create weak thermal coupling while locally disordered spins create strong thermal coupling. We also assume that heat is dissipated up to a distance $r$ (see Fig.~\ref{fig:illustration}), which can be controlled by the properties of the medium placed between the layers such as thermal conductivity and width. Thus, we assume here, that each node in one layer is thermally coupled with all the nodes in the other layer up to a distance $r$ (Fig.~\ref{fig:illustration}). In this case, the dependency of node $i$ in network $\mu$ on its interdependent nodes in network $\mu'$ (and vice versa) is reflected by the relation
\begin{equation}
    \beta_i^{\mu} = \beta \Sigma_{i'}^{\mu'}(r) \quad ,
    \label{eq: thermal_coupling}
\end{equation}
Here, $\beta = 1/ T$ is the inverse temperature, $T$, of the heat bath, $\Sigma_{i}^{\mu}(r) = \frac{1}{|K_i(r)|} \sum_{j,|i-j| \leq r} A_{ij}^\mu \sigma_j^{\mu}$ is the average magnetization of nodes within a distance $r$ from node $i$ in network $\mu$, $A_{ij}^\mu$ is the adjacency matrix of network $\mu$, and $K_i(r)$ is the set of all nodes up to a distance $r$ from node $i$ and $|K_i(r)| = \sum_{j,|i-j| \leq r} A_{ij}^\mu \sim \pi r^2$ is the number of nodes within a circle of radius $r$. Thus, locally ordered spins ($\Sigma_{i'}^{\mu'}(r) \simeq 1$) result in $\beta_i^{\mu} \simeq \beta$, i.e., the local temperature is weakly affected by the local ordering. However, as neighborhoods of spins get more and more disordered (i.e., $\Sigma_{i'}^{\mu'}(r)\to0$), the local temperatures $\beta_i^\mu\to0$, inducing a strong overheating effect. We show below that the dependency interaction range, $r$, plays a critical role and controls the propagation of thermal fluctuations. For short-range dependency, small $r$, the effect of fluctuations remains local and the ferromagnetic to paramagnetic phase transitions are continuous, while for long-range dependency, large $r$, fluctuations have global effects, and abrupt transitions are observed.
\par

\underline{\textit{Magnetic evolution.--}} The magnetic evolution of the system with time from a given initial conditions $\boldsymbol \sigma_{\mu}(0)$ can be captured using \textit{Glauber dynamics} \cite{krapivsky2010kinetic}. While taking into account the thermal coupling, a spin $\sigma_i^{\mu}$ is randomly chosen from one of the networks and will be flipped with the probability \cite{bonamassa2021interdependent}:
\begin{equation}
    \omega_i^\mu(\sigma_i^\mu) = \left( 1 + \exp \{2J \beta \sigma_i^\mu \Sigma_{i'}^{\mu'}(r) \sum_j A_{ij}^\mu \sigma_j^\mu  \} \right)^{-1}.
    \label{eq:flipping_prob}
\end{equation}
This procedure continues until the system reaches equilibrium. A single Monte Carlo step (MCS) is defined as $N$ attempts to flip randomly chosen spins and the number of steps will be considered here as time $t$ when a non-equilibrium measurement of the magnetization during a phase transition is  being conducted. In such a way, the magnetization of network $\mu$, $M_{\mu}(t) = \frac{1}{N} \sum_i \sigma_i^{\mu}(t)$, is defined using the spins configuration $\boldsymbol \sigma_{\mu}(t)$ after $t$ Monte Carlo steps from the initial conditions $\boldsymbol \sigma_{\mu}(0)$.
\par
\begin{figure*}
	\centering
	\begin{tikzpicture}[      
	every node/.style={anchor=north east,inner sep=0pt},
	x=1mm, y=1mm,
	]   
	\node (fig1) at (0,0)
	{\includegraphics[scale=0.53]{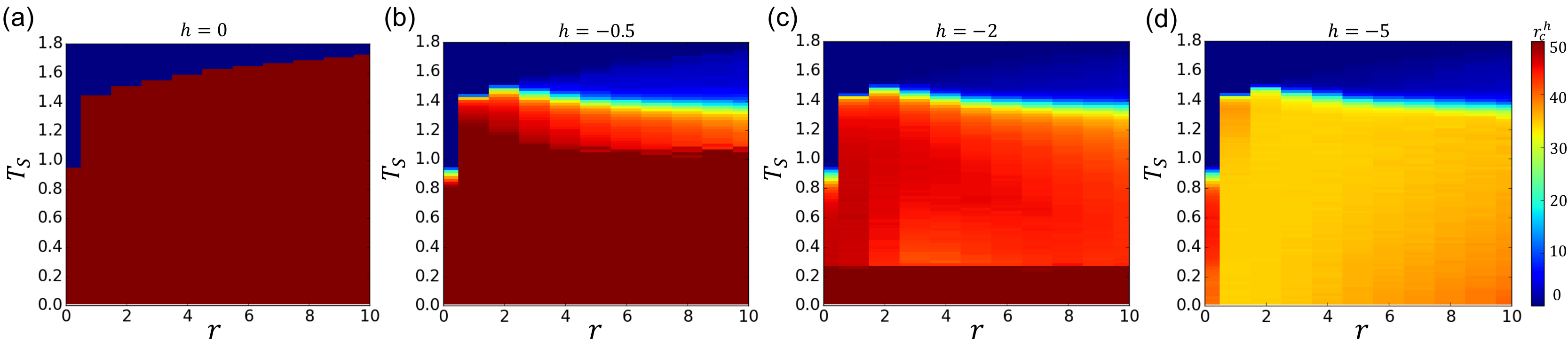}};
	\end{tikzpicture}
	\caption{\textbf{Localized magnetic field intervention. } Similar to localized heating, applying a localized negative magnetic field $h < 0$ can induce a nucleation transition. \textbf{(a)} For $h = 0$ there are no induced nucleation transitions and only two phases separated by $T_{c,<}(r)$ appear, ordered phase (brown) and disordered phase (blue). \textbf{(b) - (d)} Once a localized magnetic field is applied, a \textit{metastable regime} appears where a circular critical radius $r_c^h$ of localized field can induce a nucleation transition. The metastable regime expands with the field intensity $|h|$ towards lower temperatures, shrinking finally the ordered phase which completely disappears for large field intensity $|h| \gg 0$. Note the analogy between this PIN system studied here to percolation of abstract interdependent spatial networks~\cite{berezin2015localized}.}
	\label{fig:phase_diagram_field}	
\end{figure*}
\underline{\textit{Magnetization phase transitions.--}} We measure the steady-state magnetization as a function of temperature for different dependency range $r$ both for ordered initial conditions ($\sigma_i = +1$ for all spins in both networks) and disordered initial conditions ($\sigma_i = \pm 1$ randomly), see Fig.~\ref{fig:M_T}. Interestingly, a critical dependency range $r_c \simeq 2$ exists below which the transition is continuous similar to a single network~\cite{onsager1944crystal} but $T_{c,<}(r)$ increases with $r$, while above $r_c$ the transition becomes abrupt with hysteresis phenomena (see Fig.~\ref{fig:M_T} and inset). This behavior becomes apparent when measuring the transitions critical points $T_{c,<}(r)$ (heating) and $T_{c,>}(r)$ (cooling) from the ordered phase ($M \simeq 1$) and disordered phase ($M \simeq 0$) respectively, see inset of Fig.~\ref{fig:M_T}. The continuous transition appears for $r < r_c$ and in this case there is no hysteresis, i.e.,  $T_{c,<}(r) = T_{c,>}(r)$. In contrast, an abrupt transition and hysteresis are observed for $r > r_c$ where $T_{c,<}(r) > T_{c,>}(r)$. 
\par
While for $r > r_c$ the transition from the ordered phase to the disordered phase is abrupt, nevertheless, the transition nature depends on $r$. For low values of $r$, but still $r>r_c$, fluctuations occur locally and a spontaneous nucleation transitions are observed at $T_{c,<}(r)$. In this case, a disordered droplet is spontaneously created and radially spread in the system, see Fig.~\ref{fig:MTc_t}\textbf{(a)-(c)}. Measuring the non-equilibrium magnetization as a function of time during the transition shows a parabolic shape (Fig.~\ref{fig:MTc_t}\textbf{(a)}) as a result of the nucleation of the droplet mass is taking over (Fig.~\ref{fig:MTc_t}\textbf{(b)}). This behavior is a result of the linear increase of the droplet radius with time $R \sim t$ (Fig.~\ref{fig:MTc_t}\textbf{(c)}) resulting in the scaling of the droplet mass $M_d = \pi R^2 \sim t^2$.

\par
In contrast to the nucleation transition for low $r$ but above $r_c$, in the case of $r \gg r_c$ fluctuation has a global effect and the transition is of mixed-order \cite{boccaletti2016explosive,d2019explosive,gross2022fractal,wei-prl2012}. In this case, the non-equilibrium behavior of the magnetization at the critical point shows a \textit{plateau} similar to interdependent networks \cite{zhou2014simultaneous} where the magnetization fluctuates around a given value of the magnetization for a long time before converging into the disordered phase exponentially fast as shown in  Fig.~\ref{fig:MTc_t}\textbf{(d)}. The plateau is characterized by an almost constant number of successful flips at each time step $S_t$ shown in Fig.~\ref{fig:MTc_t}\textbf{(e)} and a critical branching factor $\eta_t = S_t/S_{t-1} \simeq 1$, see Fig.~\ref{fig:MTc_t}\textbf{(f)}.

\underline{\textit{Global macroscopic transition due to microscopic}} \underline{\textit{intervention.--}} Spatial interdependent networks can experience an induced macroscopic phase transition as a result of local microscopic interventions such as localized attacks in percolation which create a new metastable state \cite{berezin2015localized,vaknin2017spreading,vaknin2020spreading}. Here we show the physical analog of nucleation-induced transition from the ordered to the disordered phase in spatial interdependent magnetization networks both for microscopic localized heating and localized magnetic field. Localized interventions are assumed to change physical quantities such as temperature or magnetic field within a circle of a finite radius $r^h$ in one layer of the system. In the case of localized heating, spins outside the circle will experience the system temperature $T_S$ while spins within the circle also experience the localized heating $\Delta T_{L}$ with the total temperature $T =T_{S} + \Delta T_{L}$. Thus, the temperature $T_i$ felt by node $i$ can be summarized as
\begin{equation}
    T_i =\begin{cases}
			T_{S}, & \text{if $d_i > r^h$ }\\
            T_{S} + \Delta T_{L}, & \text{if $d_i \leq r^h$ }
		 \end{cases}
   \label{eq:localized_heating}
\end{equation}
where $d_i$ is the distance of node $i$ from the center of the heating. Since $T_i$ is affecting the flipping probability of spin $i$ in Eq.~\eqref{eq:flipping_prob}, spins within the circle are likely to be more disordered, and if $r^h$ is large enough, a disordered droplet 
that will spread in the system can be created and the system will experience an induced nucleation transition as demonstrated in Figs.~\ref{fig:phase_diagram_heat}\textbf{(a)-(f)}. In fact, for a given $\Delta T_{L}$, a \textit{finite} critical radius $r_c^h$ exists at a metastable regime where for any values of $T_{S}$ and $r$, for $r^h < r_c^h$ the disorder will not spread and the system will remain in the ordered phase while for $r^h > r_c^h$ the system will experience a nucleation transition into the disordered state. Note that $r^h_c$ does not depend on $L$ and therefore it is regarded as a microscopic intervention (see SI for finite size analysis). Thus, a phase diagram of $r_c^h(T_{S},r)$ can be analyzed for different heating intensities. The null case of no localized heating i.e. $\Delta T_{L} = 0$ shown in Fig.~\ref{fig:phase_diagram_heat}\textbf{(g)} display two phases, the ordered phase for $T_{S} < T_{c,<}(r)$ and the disordered phase for $T_{S} > T_{c,<}(r)$ similar to the inset of Fig.~\ref{fig:M_T}. However, once the system is localizely heated i.e. $\Delta T_{L} > 0$, a \textit{metastable} regime appears where a localized microscopic intervention of radius larger than $r_c^h$ will induce a macroscopic transition. Figs.~\ref{fig:phase_diagram_heat}\textbf{(h)-(j)} show how the metastable regime expands toward the ordered phase as the circular heating intensity increases.
\par

Similar to localized microscopic heating, applying a negative localized magnetic field $h <0$ can induce a nucleation type of macroscopic transition from the ordered phase to the disordered phase. In this case, the localized field, and the flipping probability should adjust in a similar way to Eqs.\eqref{eq:flipping_prob}-\eqref{eq:localized_heating} which can be seen in the SI. The null case of no magnetic field i.e. $h = 0$ shown in Fig.~\ref{fig:phase_diagram_field}\textbf{(a)} displays two phases similar to the null case of localized heating shown in Fig.~\ref{fig:phase_diagram_heat}\textbf{(a)}. The ordered phase is for $T_{S} < T_{c,<}(r)$ and the disordered phase for $T_{S} > T_{c,<}(r)$. However, as a localized magnetic field is applied, $h < 0$, a metastable regime appears where a finite critical microscopic radius $r_c^h$ influenced by a localized field can induce nucleation transition. The metastable regime expands as $|h|$ increases (Figs.~\ref{fig:phase_diagram_field}\textbf{(b) - (d)}) until the ordered phases completely disappear for $r > r_c$ even at zero temperature. Note that the metastable regime of the localized field expands differently compared to the localized heating in Fig.~\ref{fig:phase_diagram_heat} probably as a result of the symmetry breaking of the field. 
\par
\underline{\textit{Discussions.--}}
The extensive theoretical study of percolation of abstract interdependent networks in the last decade has led to the first-ever experiment of interdependent superconductors \cite{bonamassa2022superconductors}. The ability to perform controlled experiments in PINs is a significant breakthrough in identifying and proving  novel phase transitions. Furthermore, the study of PINs shows significantly richer phenomena compared to abstract percolation, especially in the realistic spatial case, and is expected to be a novel frontier of experimental and theoretical research of PINs. We hope our theoretical results of novel phase transitions in thermally coupled magnetic networks, and particularly the induced nucleation transition of microscopically localized interventions, will motivate experimentalists and theorists to test this theory and further study interdependent magnetic systems as well as other PINs. 
\par

\bibliographystyle{unsrt}
\bibliography{mybib}
\end{document}